\documentclass[journal]{IEEEtran}
\usepackage{cite}
\usepackage{empheq}
\usepackage{amsthm}
\usepackage{amsmath,amssymb,amsfonts,eqnarray}
\usepackage{graphicx}
\usepackage{textcomp}
\usepackage{xcolor}
\usepackage{acronym}
\usepackage{tabularx}
\usepackage[ruled]{algorithm2e}
\usepackage{algpseudocode}
\usepackage{multirow}
\usepackage{academicons}
\usepackage{scalerel}
\usepackage{tikz}
\usetikzlibrary{svg.path}
\usepackage{subcaption}
\usepackage{url}
\usepackage{ragged2e}
\usepackage{steinmetz}
\usepackage[symbol]{footmisc}
\usepackage{mathtools}
\usepackage{bm}
\usepackage{makecell}
\setcellgapes{4pt}
\usepackage{enumitem}
\usepackage{cases}
\usepackage{orcidlink}

\def\BibTeX{{\rm B\kern-.05em{\sc i\kern-.025em b}\kern-.08em
    T\kern-.1667em\lower.7ex\hbox{E}\kern-.125emX}}

\usepackage{hyperref}
 \hypersetup{backref=true,       
    pagebackref=true,               
    hyperindex=true,                
    colorlinks=true,                
    breaklinks=true,                
    urlcolor= black,                
    linkcolor= black,                
    bookmarks=true,                 
    bookmarksopen=false,
    filecolor=black,
    citecolor=black,
    linkbordercolor=black
}
\begin{document}
\title{Holographic MIMO\\ for Next Generation Non-Terrestrial Networks:\\ Motivation, Opportunities, and Challenges}
\author{Giovanni Iacovelli, Chandan Kumar Sheemar, Wali Ullah Khan, Asad Mahmood, \\George C. Alexandropoulos, Jorge Querol, and Symeon Chatzinotas
\thanks{G. Iacovelli, C. K. Sheemar, W. U. Khan, A. Mahmood, J. Querol, and S. Chatzinotas are with the Signal Processing and Communications (SIGCOM) Research Group, Interdisciplinary Centre for Security, Reliability and Trust (SnT), University of Luxembourg, 1855 Luxembourg City, Luxembourg (email: name.surname@uni.lu). G. C. Alexandropoulos is
with the Department of Informatics and Telecommunications, National and Kapodistrian University of Athens, Greece and the Department of Electrical and Computer Engineering, University of Illinois Chicago, USA (email:
alexandg@di.uoa.gr).}}

\renewcommand{\qedsymbol}{\scalebox{0.75}{$\blacksquare$}}
\newtheorem{assumption}{Assumption}
\newtheorem{remark}{Remark}
\acrodef{A2G}{Air-to-Ground}
\acrodef{AoV}{Angles of View}

\acrodef{BS}{Base Station}
\acrodef{BCD}{Block Coordinate Descendent}

\acrodef{CBR}{Constant BitRate}
\acrodef{CP}{Check-Point}

\acrodef{FoV}{Field of View}
\acrodef{FoVs}{Fields of View}

\acrodef{FR1}{Frequency Range 1}

\acrodef{GA}{Genetic Algorithm}

\acrodef{KPI}{Key Performance Index}
\acrodef{KPIs}{Key Performance Indices}

\acrodef{IoD}{Internet of Drones}
\acrodef{IoT}{Internet of Things}
\acrodef{ISAAC}{Iterative Stochastic ApproAch to constrained drones' Communications}
\acrodef{ITS}{Intelligent Transportation System}

\acrodef{LIDAR}{LIght Detection And Ranging}
\acrodef{LoS}{Line of Sight}

\acrodef{MEC}{Mobile Edge Computing}
\acrodef{MIMO}{Multiple-Input-Multiple-Output}
\acrodef{MINLP}{Mixed-Integer Non-Linear Programming}
\acrodef{MINLFP}{Mixed-Integer Non-Linear Fractional Programming}

\acrodef{NLFP}{Non-Linear Fractional Programming}
\acrodef{NOMA}{Non-Orthogonal Multiple Access}
\acrodef{NR-U}{5G NR in Unlicensed spectrum}
\acrodef{NLoS}{Non Line of Sight}

\acrodef{OoS}{Out-of-Service}

\acrodef{QoE}{Quality of Experience}

\acrodef{RGB}{Red Green and Blue}

\acrodef{SCA}{Successive Convex Approximation}
\acrodef{SoC}{System-on-chip}
\acrodef{SNR}{Signal-to-Noise Ratio}

\acrodef{UAV}{Unmanned Aerial Vehicle}
\acrodef{UAS}{Unmanned Aerial System}
\acrodef{UDP}{User Datagram Protocol}

\acrodef{VLC}{Visible Light Communications}

\acrodef{WSN}{Wireless Sensor Network}
\acrodef{IRS}{Intelligent Reflective Surface}
\acrodef{RIS}{Reconfigurable Intelligent Surface}
\acrodef{PRU}{Passive Reflective Unit}
\acrodef{GU}{Ground User}
\acrodef{OFDMA}{Orthogonal Frequency Multiple Access}
\acrodef{AtG}{Air-to-Ground}
\acrodef{AoA}{Angles of Arrival}
\acrodef{AoD}{Angles of Departure}
\acrodef{RV}{Random Variable}
\acrodef{CDF}{Cumulative Distribution Function}
\acrodef{PDF}{Probability Density Function}
\acrodef{CSI}{Channel State Information}
\acrodef{DRL}{Deep Reinforcement Learning}
\acrodef{PPO}{Proximal Policy Optimization}
\acrodef{MDP}{Markov Decision Process}
\acrodef{FBS}{Flying Base Station}
\acrodef{WPT}{Wireless Power Transfer}
\acrodef{TDMA}{Time Division Multiple Access}
\acrodef{AO}{Alternating Optimization}
\acrodef{MMSE}{Minimum Mean Squared Error}
\acrodef{SINR}{Signal-to-Interference-plus-Noise Ratio}
\acrodef{MRC}{Maximum Ratio Combining}
\acrodef{LS}{Least Squares}
\acrodef{MSE}{Mean Squared Error}
\acrodef{DFT}{Discrete Fourier Transform}
\acrodef{RMSE}{Root Mean Squared Error}
\acrodef{GEO}{Geosynchronous Equatorial Orbit}
\acrodef{LEO}{Low Earth Orbit}
\acrodef{MEO}{Medium Earth Orbit}
\acrodef{HAP}{High Altitude Platform}
\acrodef{ITU}{International Telecommunication Union}
\acrodef{NT}{Non-Terrestrial}
\acrodef{NTN}{Non-Terrestrial Network}
\acrodef{UT}{User Terminal}
\acrodef{QoS}{Quality of Service}
\acrodef{NGSO}{Non-Geostationary Orbit}
\acrodef{mmWave}{Millimeter Wave}
\acrodef{5G}{Fifth Generation}
\acrodef{6G}{Sixth Generation}
\acrodef{NR}{New Radio}
\acrodef{TDD}{Time Division Duplexing}
\acrodef{3GPP}{Third Generation Partnership Project}
\acrodef{GW}{Gateway}
\acrodef{APA}{Antenna Planar Array}
\acrodef{ASIC}{Application-Specific Integrated Circuit}
\acrodef{HMIMO}{Holographic MIMO}
\acrodef{HMIMOS}{HMIMO Surface}

\newtheorem{theorem}{Theorem}
\newtheorem{corollary}{Corollary}
\newtheorem{lemma}{Lemma}

\maketitle

\begin{abstract}
In this article, we propose the integration of the Holographic Multiple Input Multiple Output (HMIMO) as a transformative solution for next generation Non-Terrestrial Networks (NTNs), addressing key challenges, such as high hardware costs, launch expenses, and energy inefficiency. Traditional NTNs are constrained by the financial and operational limitations posed by bulky, costly antenna systems, alongside the complexities of maintaining effective communications in space. HMIMO offers a novel approach utilizing compact and lightweight arrays of densely packed radiating elements with real-time reconfiguration capabilities, thus, capable of optimizing system performance under dynamic conditions such as varying orbital dynamics and Doppler shifts. By replacing conventional antenna systems with HMIMO, the complexity and cost of satellite manufacturing and launch can be substantially reduced, enabling more streamlined and cost-effective satellite designs. This advancement holds significant potential to democratize space communications, making them accessible to a broader range of stakeholders, including smaller nations and commercial enterprises. Moreover, the inherent capabilities of HMIMO in enhancing energy efficiency, scalability, and adaptability position this technology as a key enabler of new use cases and sustainable satellite operations.
\end{abstract}


\section{Introduction}
\IEEEPARstart{W}{ireless} communications have evolved rapidly over the past few decades, propelled by the growing demand for high-speed, reliable, and widespread connectivity. Starting from the early generations of cellular networks, which primarily supported basic voice services, the wireless landscape has progressed to today’s fifth generation (5G) networks. These modern networks offer unprecedented data rates, low latency, and the ability to connect a massive number of devices simultaneously. As society increasingly relies on wireless technologies for a wide range of applications, from everyday communication to critical infrastructure, the expectations for future networks continue to rise. The next sixth generation (6G) of wireless systems seeks to push these boundaries even further by delivering ultra-reliable, high-capacity, and low-latency communication accessible everywhere, including the most remote and underserved areas \cite{wang2023road}.

The vision of ubiquitous connectivity has led to a growing interest in non-terrestrial networks (NTNs) for achieving global connectivity \cite{khan2024reconfigurable}. Such networks include unmanned aerial vehicles (UAVs), high altitude platform stations (HAPS), and satellites, such as Low Earth Orbit (LEO), Medium Earth Orbit (MEO) and Geostationary Earth Orbit (GEO) \cite{azari2022evolution}. NTNs offer the unique advantage of extending communication services to regions where traditional terrestrial networks (TNs) are either impractical or too costly to implement, such as oceans, deserts, mountainous areas, and sparsely populated regions. By bridging these connectivity gaps, NTNs are essential for providing truly global communication coverage. In particular, satellite communications have gained particular attention for their ability to offer wide-area coverage and support seamless global connectivity. Recent technological advancements, particularly in non-stationary satellite constellations, have accelerated this interest. 
\begin{figure*}[!t]
    \centering
    \includegraphics[width=0.9\linewidth]{{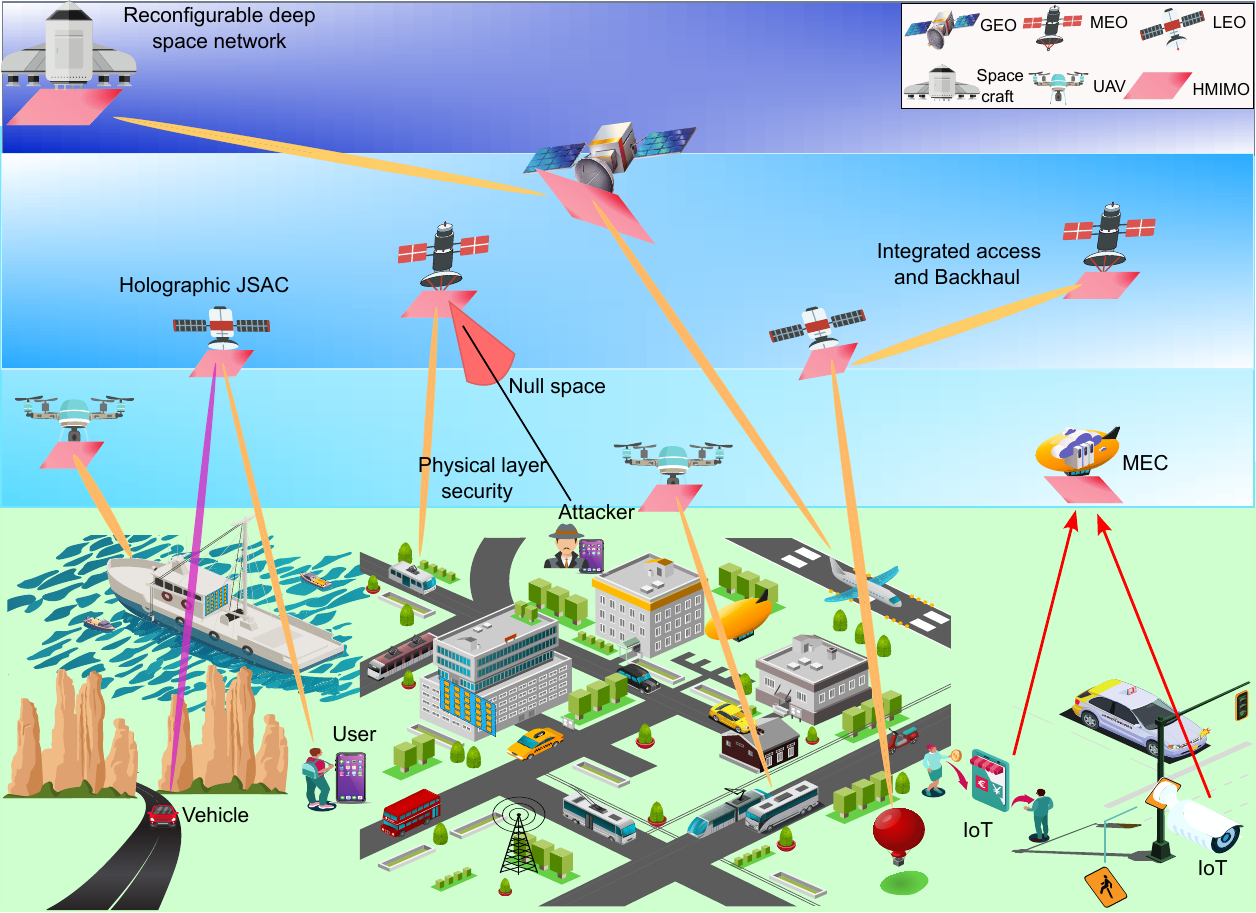}}
    \caption{HMIMO-assisted next generation NTNs. IoT stands for internet of things.}
    \label{fig:HC}
\end{figure*}

Although NTNs have the potential to revolutionize global communications, they face several significant challenges. For instance, in the case of UAVs, the limited power budget must be allocated between communication systems and trajectory control \cite{zeng2017energy}. The use of bulky antenna front-ends increases energy consumption for trajectory maintenance due to greater mass, thereby reducing overall efficiency.
In the case of satellites, manufacturing and launching costs dictated by EURO/mass remain a major concern \cite{madry2018innovative}. These expenses stem from the sophisticated design processes, the use of advanced components, and the need for large-scale, high-performance antenna arrays to support space-based massive MIMO capabilities to overcome extremely large pathloss. Furthermore, once in orbit, satellites rely on a limited power budget which must be regenerated with solar panels, making energy consumption a critical concern. Also, the high mobility of non-stationary satellites introduces additional challenges, such as Doppler shifts and rapid changes in link quality, requiring real-time reconfigurability to maintain stable communication links \cite{yeh2024efficient}. Managing interference is another significant hurdle, particularly as the number of satellites in NTNs increases. Large constellations require sophisticated coordination mechanisms to prevent signal disruption and interference management, not just within the network, but also with other satellites and terrestrial communication systems. Therefore, next-generation NTNs call for energy-efficient solutions to ensure operational longevity, small mass, and greater operational efficiency to achieve global, seamless connectivity in a sustainable fashion.

In this article, we advocate for the Holographic Multiple-Input Multi-Output (HMIMO) based new space era \cite{huang2020holographic}, which has increased potential to overcome the aforedescribed challenges of NTNs and enable sustainable space proliferation to enable new use cases. 
We shed light on how the inherent potential of HMIMO can solve the most difficult challenges of traditional NTNs, such as hardware costs, launch costs, and energy efficiency. The flexibility brought by HMIMO in NTNs systems can enable real-time reconfiguration to optimize performance under varying conditions, such as changing orbital dynamics and Doppler. By replacing expensive and bulky traditional antenna systems with compact, lightweight HMIMO arrays \cite{gong2023holographic} (see Fig.~\ref{fig:HC}), the financial barriers associated with NTNs manufacturing in general and satellites launch costs can be substantially reduced. This reduction in hardware complexity also contributes to more streamlined and cost-effective design for UAVs, HAPS, LEO, MEO, and GEO satellites, which could make advanced space communications accessible to a wider range of stakeholders, including smaller nations and commercial entities.

\section{HMIMO Preliminaries and Potential for NTNs} \label{sezione_2}

\begin{figure*}[!t]
    \centering
    \includegraphics[width=\linewidth]{{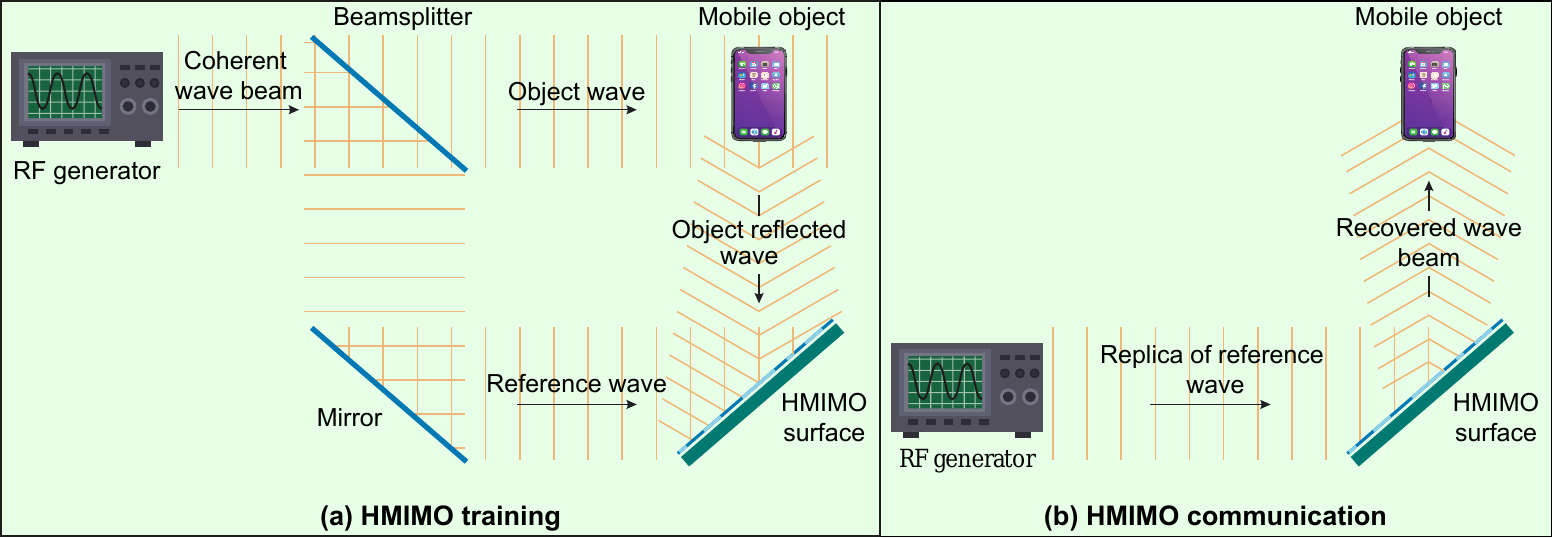}}
    \caption{The holographic principle split into two stages: (a) the HMIMO surface reconstructs the object wave by superimposing it with a reference wave; (b) once the training is complete, the HMIMO surface starts communicating with the mobile user.}
    \label{fig:HP}
\end{figure*}
\subsection{HMIMO Surfaces}
Recently, the scientific community focused on studying HMIMO as a disruptive technology to revolutionize wireless communication systems \cite{huang2020holographic}. The concepts at the basis are born as a consequence of the major advances in programmable metamaterials, already employed to realize reconfigurable intelligent surfaces (RISs) \cite{BAL2024}. An HMIMO system encompasses a massive number of reconfigurable meta-elements in a finite two-dimensional space, which, in contrast to some RIS designs, need to be placed at subwavelength spacing. Consequently, a considerable quantity of tiny and cost-effective reconfigurable elements can be systematically incorporated into a confined spatial domain, which makes the HMIMO surface quasi-continuous, thus, leading to extreme spatial multiplexing gains.  

HMIMO front-ends can serve as passive reflectors or active transceivers when equipped with radio-frequency (RF) chains \cite{BAL2024}. However, in the latter case, they only require a number of RF chains equal to the number of data streams, and do not suffer from any side lobes. The fundamental difference between an HMIMO surface and a classical Antenna Planar Array (APA) is that the latter has phase shifters, amplifiers, and attenuators at every bulky radiating element. HMIMO only has a feed line for each row of radiating elements, and each of them is equipped with a tiny and cost-effective tuning element able to efficiently modulate the capacitance, and in turn, the amplitude and phase of the signal. Thus, it is possible to perform the steering of the beam in a hologram-like way, from which the the ``Holographic Beam Forming'' term arises. This simplified and reconfigurable hardware-based design leads to a significant reduction in size, weight, cost, and power consumption \cite{gong2023holographic}, for sustainable growth of the next generation wireless systems.

The HMIMO technology is capable of achieving different kinds of wave manipulation, such as polarization and steering. The former is achieved by varying the shape of the antenna elements. The latter, instead, is realized by properly designing the wave interference patterns, which ultimately lead to multiplexing and beamforming \cite{an2023tutorial}. The functionality of the HMIMO surface relies on the holographic principle~\cite{gong2023holographic}, depicted in Fig. \ref{fig:HP}, which consists of two main stages: a) recording; and b) reconstruction. In the recording phase, a designated medium or device, such as holographic plates, captures the intensities and the phases of the hologram. This hologram is obtained through the interference pattern formed by superimposing a known reference wave with the wave scattered by the desired object. The subsequent reconstruction process involves illuminating the recording medium or device with a replica of the reference wave, achieving a precise reconstruction of the object's wave.

The sub-wavelength spacing of the HMIMO surface radiating meta-elements leads to the mutual coupling phenomenon, which calls for circuit theory to be accurately modeled. It is important to recall that mutual coupling has traditionally been viewed as detrimental in conventional communication systems, and was mitigated in massive MIMO antenna arrays by spacing adjacent antenna elements at distances of half a wavelength. However, exploiting mutual coupling in HMIMO surface can potentially enable super directivity, especially as the number of unit elements increases~\cite{9048753}. Furthermore, the quasi-continuous nature of the HMIMO surface allows the description of propagation phenomena directly in the electromagnetic (EM) domain. Consequently, the conventional Shannon theory no longer suffices to characterize HMIMO surface-assisted communications systems, and must be combined with EM wave theory. When circuit theory is also employed, the resulting paradigm undergoes the name of EM information theory~\cite{10417101, 10415512}, which represents a fundamental shift in the way of analyzing the true limits of wireless communications.

\subsection{Why HMIMO-Based NTNs?}
NTNs are set to transform global connectivity by providing unprecedented levels of coverage and reliability. As we advance towards the development of 6G technology, they must ensure sustainable resource usage, while meeting the demands for high-throughput applications. Furthermore, the upcoming emergence of highly variable data demands and services across different geographical areas drives the quest for NTNs with fast reconfigurability and flexibility. Clearly, these and future challenges cannot be adequately addressed through incremental improvements to existing techniques and hardware technologies available for NTNs. 

HMIMO-based NTNs emerge as a promising low-cost and energy-efficient solution to achieve fully reconfigurable global coverage, while significantly mitigating the challenges faced by conventional NTNs. As a matter of fact, recent prototypes have shown that HMIMO can perform fast beam switching at $100$ ns execution rate and $4$ us update rate \cite{pivotal28} with a $\pi^2$ high power gain, which results into an increased spectral efficiency of $3.30$ bits/s/Hz \cite{wang2022performances} in a Line-of-Sight (LoS) scenario. This can lead to significant gains for NTNs as they suffer extremely large pathloss compared to TNs. Moreover, when an HMIMO surface is used as an active transceiver~\cite{BAL2024}, it requires only a number of RF chains equal to the number of data streams to be transmitter and do not suffer from side lobes~\cite{gong2023holographic}. This can lead to NTNs with pencil-like beamforming, which could allow them to cover a vast geographical area by optimally channeling the limited power resources. 
Furthermore, HMIMO surfaces are built with a lightweight tuning element, which combined with subwavelength integration of the radiating meta-elements results in lightweight, small-sized, low power consumption architectures. In contrast to TNs, in the NTN context, such metrics have a major impact on the launching costs of the satellites, which can significantly influence the feasibility, duration, and economic viability of space missions. Since NTN nodes are energy-constrained, power efficiency can extend operational lifespan and reduce onboard energy usage of UAVs, HAPS, and satellites. For satellites, inefficient power utilization can significantly increase the costs and weight related to power generation and thermal management systems, jeopardizing space mission economics. In this regard, the integration of HMIMO in next generation NTNs holds promise for diminishing the necessity of hardware associated with power generation and cooling systems. This reduction in hardware requirements is anticipated to significantly contribute to further lowering the launching costs.
 
\section{Opportunities and Use Cases} \label{check}
We now elaborate on the capabilities of HMIMO surface-empowered NTNs for unlocking a new era of possibilities.  
\begin{itemize}
    \item {\bf Integrated Access and Backhaul:} Recognized by 3GPP as a cost-effective solution, IAB reduces reliance on wired backhaul links by enabling wireless backhaul through multiple hops across diverse frequency bands. The HMIMO NTNs, with its vast array of antenna elements, can enable pencil-like beams. This precision in beamforming combined with highly increased degrees of freedom will improve spectrum sharing between access and backhaul links, thus, minimizing interference and maximizing spectral efficiency. Additionally, the holographic principle enables continuous and seamless beam tracking for moving users, ensuring consistent connectivity even in challenging environments for backhual.
    \item {\bf Holographic Internet of Things:}
   Connecting IoT devices with NTNs is a promising solution as it enables reliable communication in remote areas where TNs are unavailable, or in smart cities where TNs are congested. One key challenge in enabling this connection with NTNs is the large pathloss. Specifically, since IoT devices' batteries are expected to last for decades and can only support low-power signaling, establishing a reliable connection becomes challenging. HMIMO emerges as a promising solution for IoTs, since the total received power is proportional to the antenna panel aperture. Consequently, this allows HMIMO to effectively compensate for the significant path loss experienced by low-power IoT signalling, thus, increasing received quality.  
  
\item {\bf Multi-access Edge Computing (MEC):} Integrating MEC into NTNs brings substantial benefits, especially for high-throughput applications, by enabling data offloading for complex computations due to limited on-board resources. However, it faces challenges such as long transmission delays due to large distances and limited bandwidth. HMIMO NTNs offer a promising solution to these challenges by enhancing communication throughput via the exploitation of large spatial multiplexing gain. This reduces significantly the transmission delays for greater data offloading. Additionally, HMIMO NTNs can enable direct signal processing in the EM domain, further reducing computation delays, and thereby, improving the overall performance of high-throughput NTNs.

\item {\bf Reconfigurable Deep Space Networks (DSNs):} These systems are designed to support space exploration missions by enabling long-distance communications between spacecraft and ground stations on Earth. However, these networks face significant challenges, including extreme path loss, limited bandwidth, cosmic radiation interference, and temperature fluctuations. The HMIMO technology constitutes an effective solution that can greatly improve the efficiency and capacity of DSNs by addressing these inherent challenges. To combat extreme pathloss, HMIMO-based NTNs leverage extremely large antenna arrays and advanced beamforming techniques to focus signals into narrow, precise beams, effectively mitigating signal degradation over vast distances. Further, the adaptive capabilities of HMIMO surface also allow real-time tuning of signals to compensate for environmental changes, such as temperature shifts and cosmic radiation. Additionally, the large degrees of freedom offered by HMIMO enable sophisticated spatial multiplexing, allowing multiple data streams to be sent simultaneously. This significantly increases data rates, making possible to support bandwidth-intensive applications over DSNs, such as high-definition imaging and real-time telemetry from distant space missions. 

\item \textcolor{black}{\bf Joint Communications and Sensing (JCAS):} HMIMO-based NTNs hold significant promise for enhanced JCAS. By leveraging densely packed reconfigurable antenna elements, HMIMO can simultaneously transmit and receive high-resolution signals, supporting both data communications and environmental sensing functions. This technology allows NTNs, such as satellites, UAVs and HAPS, to dynamically adjust their beam patterns for better users' channel and target tracking, optimizing the trade-off between communication throughput and sensing accuracy in real time. Furthermore, HMIMO's fine spatial resolution enhances sensing capabilities, enabling detection and tracking of objects with greater precision, even in challenging environments, like deep space or remote areas. The extensive degree of reconfigurability in HMIMO ensures better interference management, allowing the system to maintain robust communications, while accurately capturing sensing data. 

\item \textbf{Physical-layer security}: HMIMO in NTNs can enable enhanced physical-layer security mechanisms, offering a robust solution for secure and efficient communication in NTNs. The quasi-continuous aperture of HMIMO generates highly directional, pencil-like beams that are tightly focused on the intended ground users. Such precision, not only minimizes the risk of signal leakage, but also mitigates the vulnerability of communications to interception by unauthorized parties. Additionally, the large number of pencil-like beams enables optimal energy focusing, which can be exploited to create large numbers of jamming signals and null beams towards potential eavesdroppers.
\end{itemize}

\section{The Role of Densely Packed Antennas}
The main advantage of the HMIMO framework, compared to traditional MIMO systems, is that it allows to increase the electrical size of the antenna front-end without an actual increase in the physical size. Indeed, the radiative elements in APAs are located only at a half-wavelength distance, i.e., $\lambda/2$. To demonstrate the benefit brought by the closely-spaced antennas, an HMIMO NTN scenario has been simulated, including a LEO satellite equipped with an HMIMO surface communicating with a ground user in LoS conditions. The HMIMO surface was fed by a single RF chain and multiple radiative elements. The employed carrier frequency was $2.4$~GHz with a bandwidth of $20$ MHz, and the transmission power was $10$ W. The ground user was located on Earth in a direction defined by the elevation and azimuth angles corresponding to 45\textdegree and 30\textdegree, respectively. We have assumed that the Doppler effect is pre-compensated. Moreover, for the sake of ease, mutual coupling and noise spatial correlation have been neglected.

\begin{figure}
    \centering
    \includegraphics[width=\linewidth]{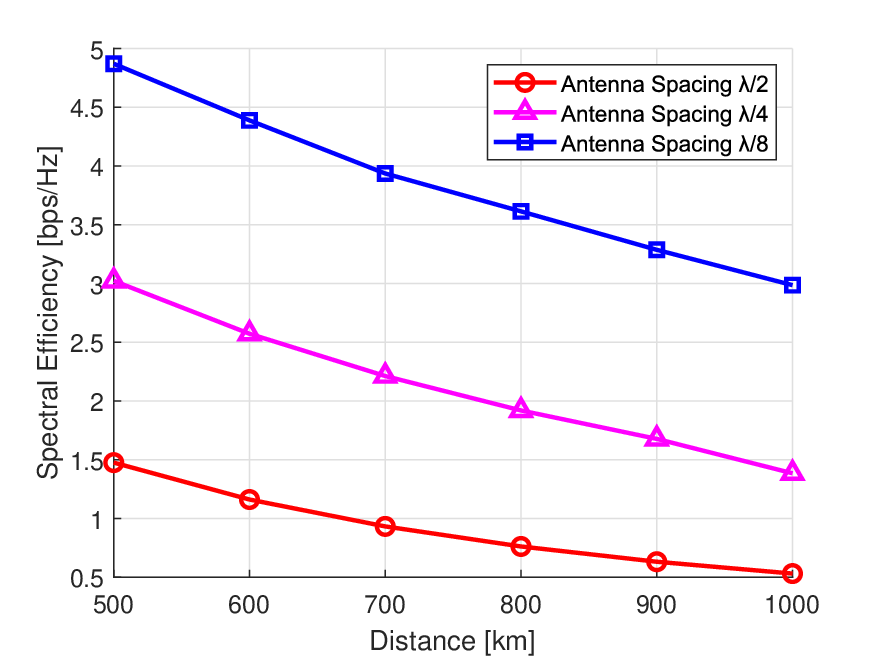}
    \caption{Achievable spectral efficiency as a function of the link distance for different antenna spacing configurations.}
    \label{fig:se}
\end{figure}

Assuming free space pathloss and perfect knowledge of channel state information (CSI) on the satellite side, a hybrid beamforming procedure was carried out to optimize the digital and analog vectors. We have specifically considered an optimization problem targeting the data rate maximization as a function of the beamforming vectors, subject to a power constraint. Figure~\ref{fig:se} shows the achievable spectral efficiency when the distance and antenna spacing vary. Specifically, the distance spanned from $500$ to $1000$ km, while three configurations were chosen for the antenna spacing: a traditional MIMO configuration with $\lambda/2$ and two HMIMO ones with $\lambda/4$ and $\lambda/8$. As expected, the latter two provide a remarkable gain in spectral efficiency. In particular, a spacing of $\lambda/8$ yields spectral efficiency levels that are $3$ to $6$ times higher with respect to traditional MIMO, i.e., $\lambda/2$ antenna spacing, for the considered parameter settings.

\section{Challenges and Future Research Directions} \label{risultati}
\begin{figure}
    \centering
    \includegraphics[width=.8\linewidth]{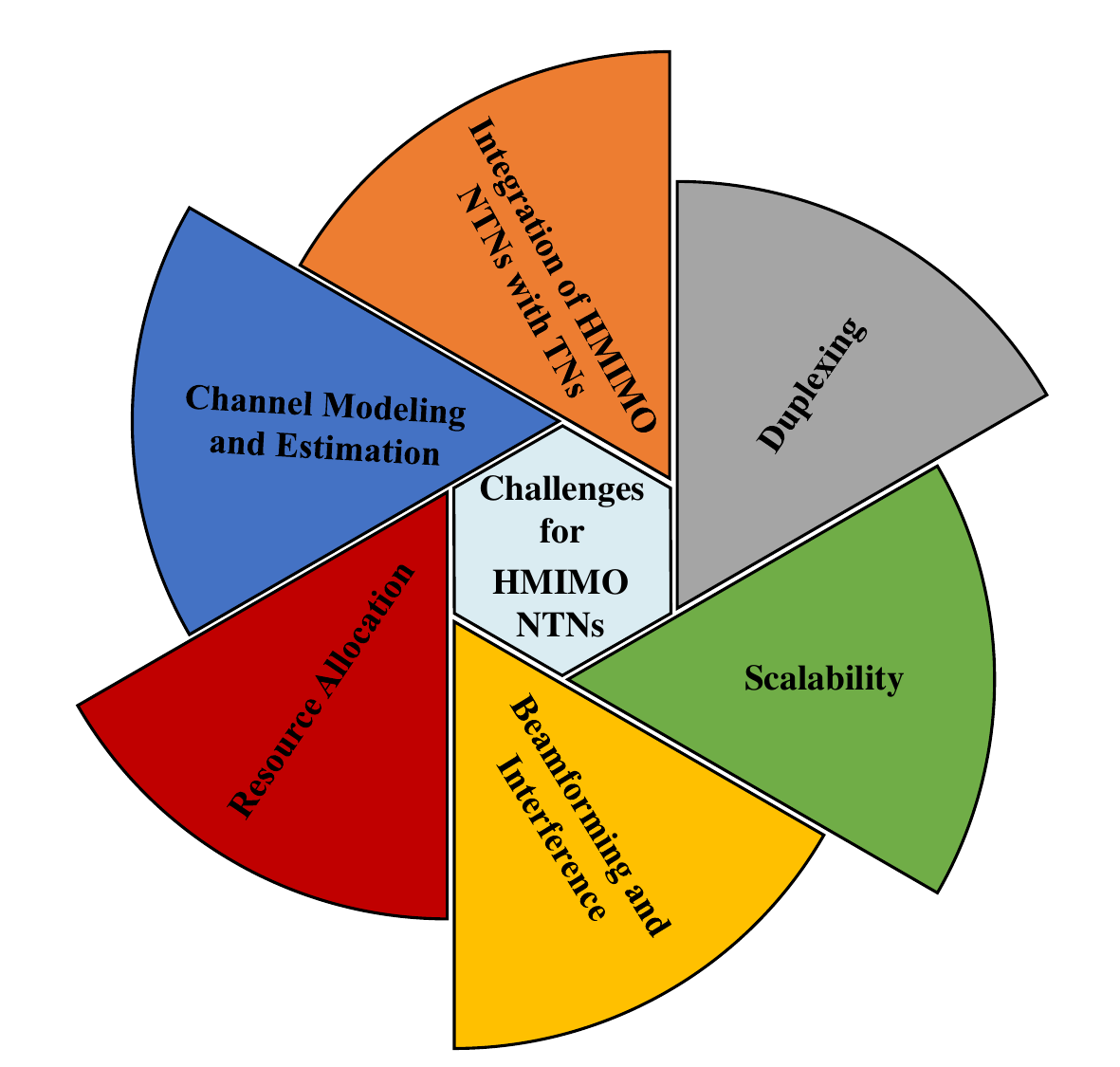}
    \caption{Key Challenges for HMIMO-based NTNs.}
    \label{fig:enter-label}
\end{figure}

This section presents key challeges that need to be addressed to enable HMIMO-based NTNs space era (see Fig.~\ref{fig:enter-label}), together with guidelines for future research directions to address them.

\textbf{Channel Modeling and Estimation}: 
Recent research has begun to explore CSI acquisition for HMIMO systems, although primarily focusing on TN scenarios. The key distinction between HMIMO and traditional MIMO communications lies in the spatial correlation and mutual coupling effects, which are exacerbated by the sub-half-wavelength element spacing in HMIMO arrays. On top of that, in the NTN context, additional propagation phenomena must be considered, such as non-LoS paths, Doppler shifts, large propagation delays, atmospheric fading, and cosmic radiation. These factors make existing MIMO channel models and estimation techniques unsuitable for direct application to HMIMO-based NTNs, necessitating significant customization. Indeed, the huge number of elements can drastically increase the estimation complexity, not only in terms of computational demand, but also in terms of time convergence due to the higher number of parameters involved. Henceforth, accurate low complexity channel estimation and tracking algorithms are necessary to decrease the computational overhead, especially given the high mobility that characterizes NTNs.

\textbf{Resource Allocation}: 
HMIMO-based NTNs are affected by mutual coupling and correlation, which can ultimately hinder communication performance if not taken into account. Furthermore, for non-stationary NTNs such as UAVs, HAPS, and LEO satellites, additional challenges arise due to their mobility, including faster channel variations and Doppler shifts. As a result, resource allocation strategies must be designed to account for these factors as well, together with the large propagation delays due to distances. The associated large-scale optimization problems for HMIMO-based NTNs will be characterized by search spaces that can vary significantly. This necessitates the development of low-complexity adaptive strategies capable of meeting the dynamic nature of the environment while minimizing latency and ensuring robust system performance under varying dynamic channel conditions. Additionally, these strategies must be scalable and efficient to accommodate the growing demands of future NTNs with large user densities and with limited on board resources.

\textbf{Beamforming and Interference}: HMIMO-based NTNs offer significant advantages in terms of beamforming capabilities, primarily due to to the HMIMO quasi-continuous aperture and the large numbers of spatial degrees of freedom. 
However, the vast majority of the solutions proposed in literature do not take into account several important aspects that can undermine the true potential of HMIMO beamforming for NTNs. The beamforming strategies, especially in the analog/holographic domain, can face an unprecedented computation complexity due to the dimension of the search space and the real constraints coming with the amplitude control, which must be met with the limited computational resources on board for fastly varying dynamic channel conditions. Indeed, interference management requires the estimation of way more parameters than the traditional MIMO systems, to align the beams. In addition, the large propagation delays inherent in NTNs cause CSI to become outdated, complicating the beamforming process. Consequently, such challenges must be considered to enable holographic beamforming on board.

\textbf{Scalability:} As pointed out earlier in the context of resource allocation and beamforming/interference management for HMIMO-based NTNs, the large number of radiating elements could potentially increase overall system complexity, especially in discrete HMIMO surface where computing direct integrals might be challenging, hindering scalability. Furthermore, given the limited onboard resources, the communication overhead required to coordinate various communication strategies between NTNs on different layers, such as HAPS, UAVs and satellites, might be significant. Therefore, to develop large-scale HMIMO-based NTNs, such as mega-constellations for achieving global coverage, low-complexity strategies with minimal communication overhead that can be executed independently at each NTN node must be developed.

\textbf{Duplexing:} For large antenna systems, including HMIMO, time division duplexing (TDD) is preferred over frequency division duplexing (FDD), primarily due to channel reciprocity. This allows the base station to estimate the downlink channel based on uplink measurements, significantly reducing the need for feedback, unlike FDD, which requires separate channel estimation for different frequencies. As the number of antennas increases, TDD also offers better scalability by minimizing feedback overhead and reducing complexity, making it more efficient for massive MIMO applications. This makes the HMIMO easible for the UAVs and HAPS. However, traditional satellite systems are designed around FDD mainly to overcome the challenge of large propagation delays, which could make CSI obsolete or inaccurate by the time it is switched between uplink and downlink modes. Consequently, novel strategies based on FDD which could make HMIMO easily adaptable in the satellite systems need to be studied.

\textbf{Integration of HMIMO-based NTNs with TNs:} HMIMO offers the potential to revolutionize the integration of TNs and NTNs, facilitating smoother communication between ground-based systems and mobile platforms, like satellites or UAVs. 
The major challenge lies in the propagation delay, particularly in NTNs with satellites in high orbits, such as GEO, where the vast distance the signal has to travel results in significantly higher latency compared to TNs, thus, severely impacting time-sensitive applications.
Furthermore, the communication overhead for coordination between TNs and NTNs is increased due to the need for precise control signaling for CSE estimation, beam alignment, and synchronization. This additional overhead can place a strain on network capacity, as real-time information exchange is critical for maintaining efficient operation across the combined systems. These challenges must be addressed for HMIMO-based NTNs to fully realize its potential in mitigating latency and ensuring seamless integration with the next generation of TNs.

\section{Conclusion} \label{conc}
In this work, we proposed the integration of the HMIMO technology for transforming next generation NTNs. This innovative antenna/metasurface technology approach leverages compact, lightweight antenna arrays with real-time reconfiguration capabilities, enabling the optimization of system performance even under the dynamic conditions typical of NTNs, such as variations in orbital dynamics and Doppler shifts. By incorporating HMIMO, NTNs can support a broader range of use cases, taking advantage of their unique characteristics. As this technology evolves, new challenges continue to arise. We have explored these by examining essential factors that must be addressed to make HMIMO-based NTNs a reality.

\bibliographystyle{IEEEtran}
\bibliography{HMIMO_NTNs}

\end{document}